\documentclass[]{aa}
\usepackage{txfonts}

\usepackage{graphicx}

\begin{document}

\title{Accretion mode changes in Centaurus X-3}

\author{B. Paul$^{1}$, H. Raichur$^{1,2}$, and U. Mukherjee$^{1}$}

\offprints{B. Paul}

\institute{$^{1}$Tata Institute of Fundamental Research, Homi Bhabha Road, Colaba,
Mumbai 400005, India\\
$^{2}$Joint Astronomy Programme, Indian Institute of Science, Bangalore, 560012 India\\
\email{bpaul@tifr.res.in}}

\date{}

\abstract{
We report here discovery of the existence of two different accretion modes in
the high mass X-ray binary pulsar Cen X-3 during its high states.
The multiband X-ray light curves of Cen X-3 lasting for more than 3400 days
obtained with the All Sky Monitor (ASM) onboard the Rossi X-ray Timing Explorer
(RXTE) shows many episodes of high and low X-ray intensities.
The high intensity phases last between a few to upto 110 days and the
separation between two high intensity phases also varies widely.
One remarkable feature deduced from the RXTE-ASM light curves is that during
these high intensity phases, Cen X-3 manifests in two very distinct spectral
states.
When the source makes a transition from the low intensity phase to the high
intensity phase, it adopts one of these two spectral states and during the
entire high intensity phase remains in that particular spectral state.
During December 2000 to April 2004, all the high intensity episodes
showed a hardness ratio which is significantly larger than the same during
all the high states prior to and subsequent to this period.
It is also found that most of the soft outbursts reach a nearly constant peak
flux in the 5--12 keV band.
For comparison, similar analysis was carried out on the long term X-ray light
curves of three other X-ray binary pulsars Her X-1, Vela X-1, and SMC X-1.
Results obtained with these sources are also presented here and we found
that none of the other sources show such a behaviour.
From these observations, we suggest that Cen X-3 has two different accretion
modes and in the course of nine years it has exhibited two switch overs between
these.

\keywords{stars : neutron --- Pulsars : individual (Cen X-3) --- X-rays : stars}

}

\maketitle

\section{Introduction}

The accretion powered X-ray Binary Pulsars (XBP) show X-ray intensity modulations
related to the spin of the neutron star, orbital period of the binary, and
quasi-periodic oscillations related to motion in the accretion disk.
Apart from these, many of the sources also show other short and long time
scale intensity variations (Psaltis 2005).
Periodic or semi-periodic long term intensity variations are known to be
present in a few systems like Her X-1, LMC X-4 and SMC X-1 (Clarkson et al.
2003b, Boyd et al. 2004) which may have a different origin compared to the
long term intensity variations in other low and high mass X-ray binaries
(Paul, Kitamoto \& Makino 2000; Clarkson et al. 2003b;  Still \& Boyd 2004).
The quasi-periodic long term intensity variations observed in many X-ray
binaries can be due to the precession of a warped inner accretion disk
(Shirakawa \& Lai 2002).
A stable superorbital period of the warped disk is more common in the high
magnetic field neutron star systems (Oglivie \& Dubus 2001).
Often the intensity variations at the superorbital period is found to be
associated with corresponding complex spectral changes (Naik \& Paul 2002).
The spectral changes can be due to variations in the absorption column density
in the line of sight and can also have very different fluorescence iron line flux
and equivalent width (Naik \& Paul 2004).
Long term intensity variations, observed in many other binary systems 
with low and high mass companion stars are believed to be due to variations in
the mass accretion rate caused by changes in the accretion disk characteristics
or activity of the companion star (Paul et al. 2000).

Here, we report our investigation of the coarse spectral characteristics of
a very well studied High Mass X-ray Binary (HMXB) pulsar Cen X-3
with multi-band light curves taken with the All Sky Monitor (ASM) of the Rossi
X-ray Timing Explorer (RXTE) over a long period of 3400 days.
A comparison is also made with three other accretion powered X-ray pulsars
Her X-1, Vela X-1, and SMC X-1.

\begin{figure*}[t]
\begin{center}
\includegraphics[angle=-90,width=5.8in]{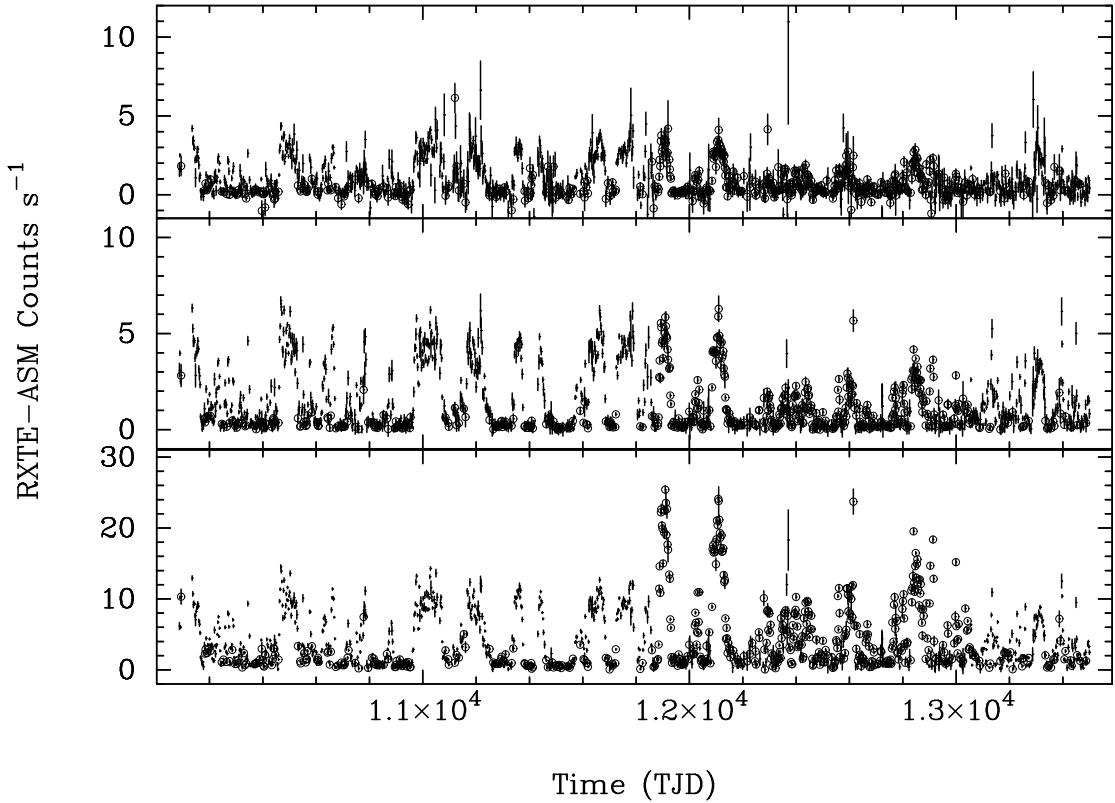}
\end{center}
\caption
{RXTE-ASM light curves of Cen X-3
in three different energy bands of 1.5--3 keV, 3--5 keV and
5--12 keV are shown here in top, middle, and bottom panels respectively.
Circles are used for data points with hardness ratio (between 5--12 keV and
3--5 keV count rates) greater than 3.5.}
\end{figure*}

\section{Observations, Analysis and Results}

The RXTE-ASM detectors scan the sky in a series of 90 s dwells in three
energy bands, 1.5--3, 3--5, and 5--12 keV.
For details of the instrument and its performance in space, refer to Levine
et al. (1996).
Here we have used the multi band, per-dwell light curves of
Cen X-3 and three other bright XBPs Her X-1, Vela X-1, and SMC X-1
provided by the RXTE-ASM team,
covering the period of 5 January, 1996 to 12 May, 2005.

The light curves of Cen X-3, in the three different energy bands of ASM,
are shown in Figure 1.
As described later, this figure was generated after removing data obtained
during the X-ray eclipse of Cen X-3 and is binned with a period equal to
the orbital period of the binary system.
It is clear that the X-ray luminosity of Cen X-3 has a wide variability. 
The source frequently switches between low and high states with
intensity varying by a large factor.
In all the three energy bands the peak flux during the bursts is larger
compared to the low state flux by upto a factor of 40.
The low and high intensity states last between a few to upto 110 days
and occur with no apparent periodicity.
The X-ray flux during the low states averaged over a few days is always
found to be within about 30\% of the average low state flux during the
entire observation period.
Period searches with different algorithms did not show any long term
periodicity in the intensity variation of Cen X-3.
The superorbital intensity variations in two other HMXB pulsars, LMC X-4
(Paul \& Kitamoto 2002) and SMC X-1 (Clarkson et al. 2003a)
are smooth in nature, almost sinusoidal over the superorbital period.
The other X-ray binary pulsar that shows prominent and
stable superorbital intensity variations, when it is out of the
occasional anomalous low states is Her X-1 (Still \& Boyd 2004).
In this source, each superorbital period consists of one main-on phase and one
short-on phase separated by two low intensity phases.
In the case of Cen X-3, the rise and fall in intensity around the high
states is quite abrupt, lasting about 2--8 days.
One interesting feature of the light curve prior to December 2000 (TJD 11880)
is that there appears to be a ceiling in the maximum luminosity that the source 
reaches during the outbursts. The ceiling is most clearly visible in the 5--12
keV band and seems to be present in the 3--5 keV band as well.

For the purpose of this work, we examined the coarse spectral evolution of
Cen X-3 as the source went through its intensity changes and calculated two
hardness ratios HR1 (3--5 keV/1.5--3 keV) and HR2 (5--12 keV/3--5 keV) from
the RXTE-ASM data.
A plot of HR1 and HR2 against the ASM count rate is shown in Figure 2.
During the low states, the hardness ratio measurements have large
error bars, and in Figure 2 we have plotted only the data points during
which the average ASM count rate is more than 2.
The hardness ratio measured during the high states clearly 
shows a change in spectral shape of the source in the period between
December 2000 (TJD 11800) and April 2004 (TJD 13100).
A plot of the medium energy (3--5 keV) X-ray count rate against the hard X-ray
(5--12 keV) count rate very clearly shows the existence of two different spectral
states in Cen X-3 (Figure 3).
The data points belonging to the two arms in Figure 3 are from different
time ranges in the light curves and are marked in Figure 1 according to the
hardness ratio.
Below we describe a careful examination of the various possibilities that may
produce such hardness ratio changes artificially.
We have ruled out any possibility of the observed spectral
changes being an artifact of instrument calibration, sampling etc.

\begin{figure}[t]
\begin{center}
\includegraphics[angle=-90,width=3in]{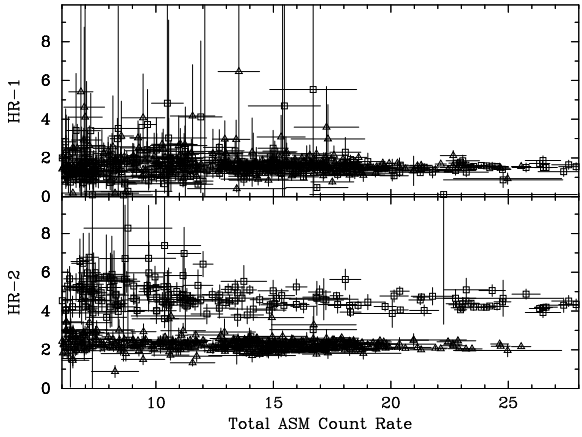}
\end{center}
\caption
{The two hardness ratios HR-1 and HR-2 derived from the RXTE-ASM light curves
of Cen X-3 are plotted here against the ASM count rate for value greater than
2 per second. Data obtained between TJD 11800 and 13100 are shown with square
symbol.}
\end{figure}

\begin{table}[b]
\caption{Orbital parameters of the four XBP sources derived from RXTE-ASM light curves}
\label{table:1}
\centering    
\renewcommand{\footnoterule}{}
\begin{tabular}{l l l l} 
\hline\hline            
Source & Orbital & Mid-eclipse  & Eclipse duration \\ 
name & period (day) & time (TJD) & in orbital phase with \\ 
     &              &            & ingress and egress\\
\hline                      
   Cen X-3$^{\mathrm{a}}$  & 2.08702(3) & 10087.295(1) & 0.306 \\
   Her X-1$^{\mathrm{b}}$  & 1.700165(12) & 10088.236(3) & 0.318 \\
   Vela X-1$^{\mathrm{c}}$ & 3.89201(11) & 10087.286(6) & 0.231 \\
   SMC X-1$^{\mathrm{d}}$  & 8.96461(35) & 10087.542(4) & 0.377 \\
\hline                    
\end{tabular}
\begin{list}{}{}
\item
{\tiny
For orbital parameters of these sources determined from pulse arrival
time delays, see
$^{\mathrm{a}}$Nagase et al. 1992,
$^{\mathrm{b}}$Bhatt et al. 2003,
$^{\mathrm{c}}$Deeter et al. 1987, and
$^{\mathrm{d}}$Wojdowski et al. 1998
}
\end{list}
\end{table}

Since this binary system has a small orbital period of 2.09 day, a high mass
($\sim 20 M_{\sun}$) companion star, and an inclination angle of 70$^\circ$
(Ash et al. 1999), it shows a long eclipse lasting for about one fourth of the
orbital period during which the X-ray intensity decreases by a large factor.
The residual X-ray emission seen during the eclipse is due to the reprocessing
of the pulsar X-rays by the stellar wind of the companion star.
For individual point X-ray sources, the RXTE-ASM detectors have very short
exposures of about 90 s several times a day.
Since eclipse data would contaminate the average X-ray flux measurement,
all the figures were generated after removing the eclipse data.
For all the four sources, the orbital period, mid-eclipse time, and eclipse
duration were first derived from the unfiltered RXTE-ASM light curves and are
given in Table 1.
The orbital periods, derived from the RXTE-ASM data are consistent with the
same derived from pulse arrival time analysis.
Due to asymmetric nature of the X-ray eclipses, the mid-eclipse times derived
from X-ray light curves are not accurate and
should not be combined with pulse timing data for study of orbital period evolution.
The eclipse durations given in Table 1, include ingress and the egress.
To avoid the effect of intensity variations with orbital phase, all the
light curves and hardness ratios presented here were produced with a bin size
equal to or twice the orbital period.

The RXTE-ASM has three independent detectors with different
viewing orientations and
makes intensity image of a large part of the sky with a few arc minute
angular resolution in one rotation around its axis.
Due to this arrangement, the three detectors may obtain different effective
exposures for any X-ray source.
One of the three detectors is known to have a slightly different energy
calibration (A. Levine: private communication).
This fact, combined with the varying exposure may also produce some artificial
hardness ratio changes in the light curves.
To examine this in detail, we have separately analysed data from the three
detectors of RXTE-ASM and same result as presented in Figures 2 \& 3 were
obtained from all the three detectors.

\begin{figure}[t]
\begin{center}
\includegraphics[angle=-90,width=3in]{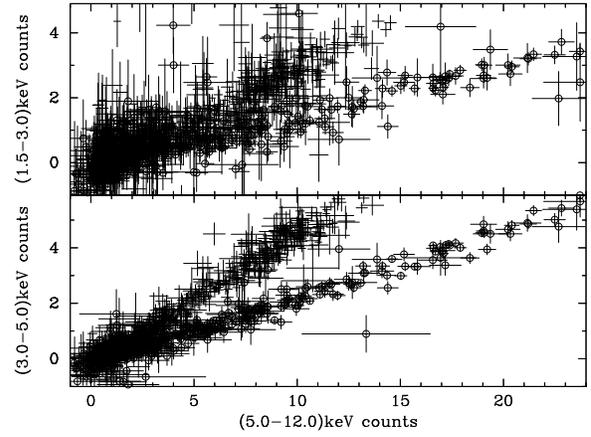}
\end{center}
\caption
{The RXTE-ASM count rates of Cen X-3 in soft (1.5--3 keV) and medium 
(3--5 keV) energy bands are plotted against the hard (5--12 keV) X-ray
count rate.
Data obtained between TJD 11800 and 13100 are shown as circles.}
\end{figure}	 

\begin{figure*}[t]
\begin{center}
\includegraphics[angle=-90,width=5.8in]{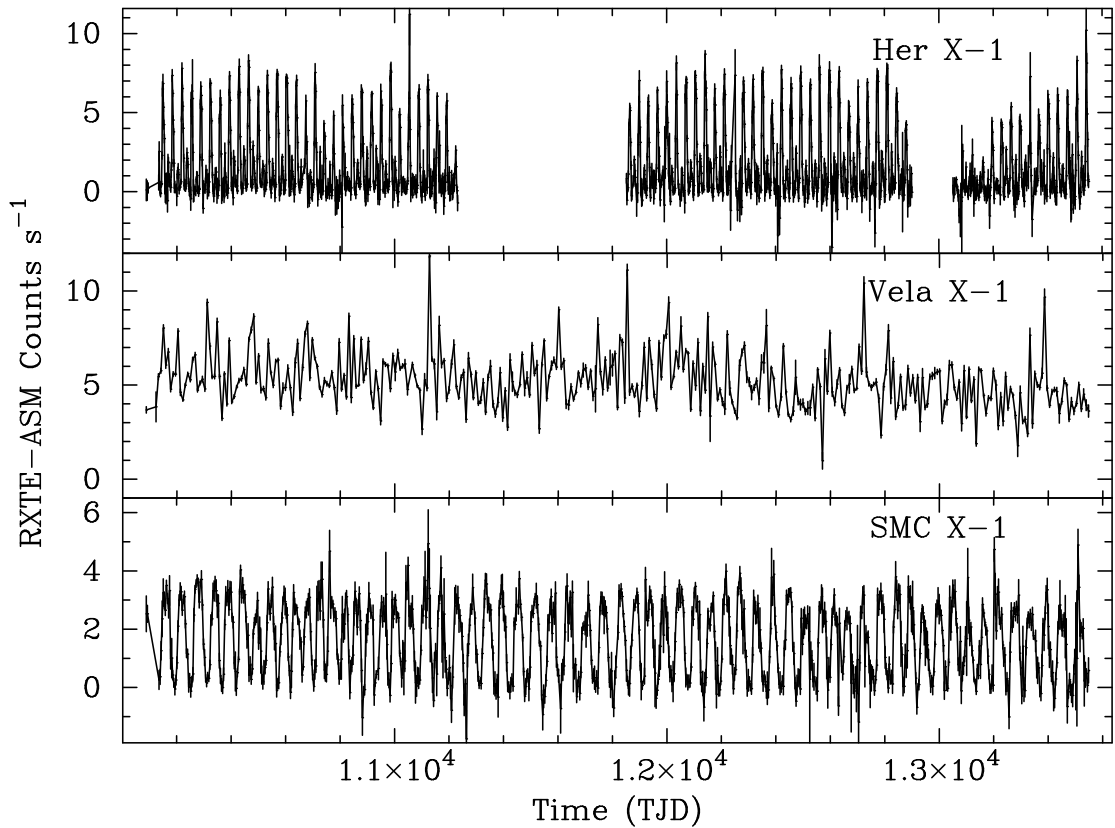}
\end{center}
\caption
{The 1.5--12 keV RXTE-ASM light curves of Her X-1, Vela X-1, and SMC X-1 are
plotted in the top, middle, and bottom panels respectively.
For Her X-1, data obtained during the two anomalous low states were omitted.}
\end{figure*}	 

\begin{figure*}[t]
\begin{center}
\includegraphics[angle=-90,width=6.0in]{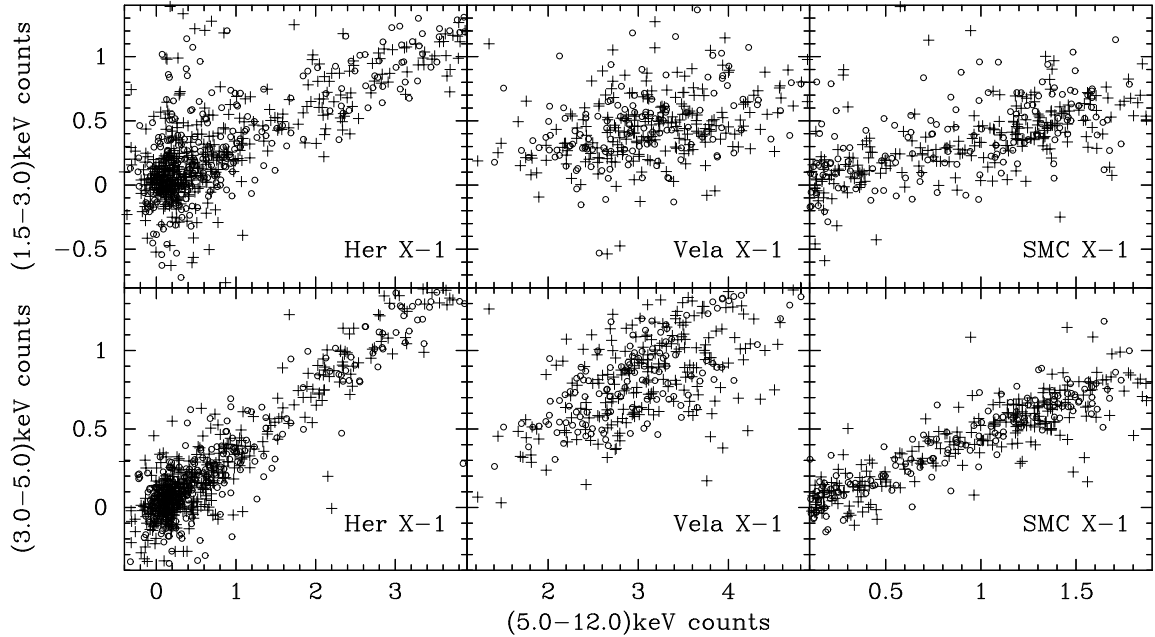}
\end{center}
\caption
{The RXTE-ASM count rates in soft (1.5--3 keV) and medium (3--5 keV) energy
bands are plotted against the hard (5--12 keV) X-ray count rates for the three
sources Her X-1, Vela X-1 and SMC X-1.
As in Figure 3, data obtained between TJD 11800
and 13100 are shown as circles. The error bars have not been shown for the sake
of clarity.}
\end{figure*}

To ensure that the observed change in spectral hardness ratio for the period
December 2000 to April 2004 is not an artifact of changing energy gain calibration
of the detectors, we have carried out the same extensive analysis for three
other XBP sources Her X-1, Vela X-1 and SMC X-1.
As was done with Cen X-3, the data acquired during the eclipses were excluded
before generating the light curves.
Light curves of these three bright XBPs in the broad energy bands of 1.5-12 keV
are shown in Figure 4 with a binsize equal to the respective orbital periods.
The duration of the light curves of these three XBPs (Figure 4) is same as that
of Cen X-3 (Figure 1).
In the case of Her X-1, data collected during the two anomalous low states 
(Boyd et al. 2004) were also excluded from analysis.
In Figure 5, the low energy (1.5--3 keV) and medium energy (3--5 keV) X-ray
count rates are plotted against the hard X-ray (5--12 keV) count rate for the
three sources Her X-1, Vela X-1 and SMC X-1.
An integration time twice the orbital period was used for Her X-1 and
SMC X-1, while for Vela X-1, it is same as the orbital period.
Data obtained between TJD 11800 and 13100 are shown with different symbols.
From Figure 5, one can clearly rule out the possibility that the changes in the
hardness ratio detected in Cen X-3 between December 2000
to April 2004 are produced artificially due to any change
in energy gain calibration of the detector system.
Count rates in the medium (3--5 keV) and hard X-ray (5--12 keV) appear to be
very well correlated for Her X-1 and SMC X-1, while for Vela X-1 there is
significant scatter.
We found Cen X-3 to be the only source which had a change in spectral hardness
ratio during the period mentioned above.

\section{Discussion}

The continuum energy spectrum of XBP sources is often described by a power-law
modified by absorption at low energies, an exponential cutoff at high energies,
and cyclotron absorption resonance features (Nagase 1989).
Intensity variations in both transient and persistent XBPs are often associated
with changes in the X-ray spectrum and pulse shape, presumably due to
modifications in the accretion column near the magnetic poles.
Study of the X-ray spectrum in different intensity states is therefore
important to understand the reason behind the long term intensity variations.

In the high state, Cen X-3 is the brightest accreting X-ray pulsar and one of the
brightest X-ray sources in the energy band of RXTE-ASM.
In the present work we have clearly established a few aspects of the long term
spectral characteristics of Cen X-3, especially in its high states.
In the high state, it has two distinct spectral modes and during each outburst
the source goes through flux changes via only one of the two spectral modes.
During any of the individual high states the source never switches between the
two spectral modes. 
There is no significant spectral evolution during each outburst, at least 
above a flux level of 100 mCrab.
All outbursts belonging to one of the two spectral modes have spectral hardness
ratio in a very narrow range (Figures 2 \& 3).

Lack of spectral variation during the individual high states is a character
of Cen X-3
that is different from sources with precessing warped inner accretion disk.
For example, Her X-1 shows a large variation in absorption column density over
its 35 day super-orbital period (Naik \& Paul 2003).
Presence of the two spectral modes is most clearly seen in the HR-2 (Figure 2)
and is less obvious in HR-1.
This indicates that the harder spectrum that occurred during December 2000 to
April 2004 is not due to a change in the absorption column density.
A change in column density in this period should have had greater effect in the
lowest energy band causing a larger difference in the corresponding HR-1 value.

One intriguing feature of Cen X-3 is a ceiling in the maximum luminosity in outbursts
prior to the first change in spectral mode that took place around TJD 11880 (Fig 1,
bottom panel).
We may speculate that it is related to the Eddington luminosity in Cen X-3 for
some limited solid angle in the soft spectral mode.

The pulsed hard X-ray (20--50 keV) flux history of Cen X-3,
obtained with CGRO-BATSE shows long term intensity variations
identical to  variations in the soft X-rays, at least for the
period of overlap between CGRO and RXTE observatories (January 1996 to May 2000).
However, since the accretion mode with harder spectral shape started after the
CGRO era, it is not possible to carry out the corresponding study involving
hard X-ray photon flux.

A detailed analysis of the X-ray spectrum in different intensity states of Cen X-3
is in progress and will be reported separately.

\begin{acknowledgements}

We thank an anonymous referee for valuable suggestions to improve the paper.
We thank the RXTE-ASM and CGRO-BATSE team for providing the valuable data.
This research has made use of data obtained from the High Energy Astrophysics
Science Archive Research Center (HEASARC), provided by NASA's Goddard Space
Flight Center.

\end{acknowledgements}

\end{document}